\documentclass{llncs}

\usepackage[utf8x]{inputenc}
\usepackage[hyphens]{url}
\usepackage[hidelinks]{hyperref}
\hypersetup{breaklinks=true}
\usepackage{subcaption}
\usepackage{longtable}
\usepackage[binary-units]{siunitx}
\usepackage[T1]{fontenc}
\usepackage{cite}
\usepackage{amsmath}
\usepackage{xcolor}
\usepackage{listings}
\usepackage{lmodern}
\usepackage{subcaption}
\usepackage{times}

\usepackage{tikz}
\usetikzlibrary{shapes}
\usetikzlibrary{arrows}

\begin{document}

\pagestyle{headings}

\title{Translating Scala Programs to Isabelle/HOL}
\subtitle{System Description}
\titlerunning{Translating Scala Programs to Isabelle/HOL}

\author{Lars Hupel\inst{1} \and Viktor Kuncak\inst{2}}
\authorrunning{L. Hupel and V. Kuncak}

\institute{Technische Universität München \and École Polytechnique Fédérale de Lausanne (EPFL)}

\mainmatter
\maketitle

\definecolor{kwgreen}{RGB}{0,128,0}
\definecolor{stringred}{RGB}{186,33,33}

\newcommand\isacommand[1]{\texttt{\textbf{\color{kwgreen}#1}}}
\newcommand{\purescala}{Pure Scala}

\lstset{
  basicstyle=\ttfamily\small,
  keywordstyle=\bfseries\color{kwgreen},
  stringstyle=\color{stringred},
  showstringspaces=false
}

\lstdefinelanguage{Isar}[]{ML}{
  morekeywords={lemma,where,by}
}

\lstdefinelanguage{Scala}{
  alsoletter={@,=,>},
  morekeywords={abstract, case, catch, choose, class, def, do, else, extends, false, final, finally, for, if, implicit, import, match, new, null, object, lazy, let,
override, package, private, protected, requires, return, sealed, super, this, throw, trait, try, true, type, val, var, while, with, yield, domain, 
postcondition, precondition, constraint, assert, forAll, _, return, @generator, ensure, require, ensuring, assuming, otherwise, asserting, holds}
  sensitive=true,
  morecomment=[l]{//},
  morecomment=[s]{/*}{*/},
  morestring=[b]",
  morestring=[b]"""
}

\begin{abstract}
  We present a trustworthy connection between the Leon verification system and the Isabelle proof assistant.
  Leon is a system for verifying functional Scala programs.
  It uses a variety of automated theorem provers (ATPs) to check verification conditions (VCs) stemming from the input program.
  Isabelle, on the other hand, is an interactive theorem prover used to verify mathematical specifications using its own input language Isabelle/Isar.
  Users specify (inductive) definitions and write proofs about them manually, albeit with the help of semi-automated tactics.
  The integration of these two systems allows us to exploit Isabelle's rich standard library and give greater confidence guarantees in the correctness of analysed programs.

\end{abstract}

\section{Introduction}

This system description presents a new tool that aims to
connect two important worlds: the world of interactive proof
assistant users who create a body of verified theorems,
and the world of professional programmers who increasingly adopt
functional programming to develop important applications.  The Scala 
language (\url{www.scala-lang.org}) enjoys a prominent role today for its
adoption in industry, a trend
most recently driven by the Apache Spark data analysis
framework (to which, e.g., IBM committed 3500 researchers recently \cite{terdoslavich2015spark}).
We hope to introduce some of the many Scala users to formal methods by providing tools they
can use directly on Scala code.
Leon system (\url{http://leon.epfl.ch}) is a verification and synthesis system for a subset of Scala 
\cite{blanc2013overview, kuncak2015leon}. Leon reuses the Scala compiler's parsing and type-checking frontend and subsequently derives verification conditions to be 
solved by the automated theorem provers, such as \emph{Z3} \cite{demoura2008z3} and \emph{CVC4} \cite{barrett2011cvc4}.
 Some of these conditions arise naturally upon use of particular Scala language constructs (e.g.\ completeness for pattern matching), whereas others stem 
from Scala assertions (\verb~require~ and \verb~ensuring~) and can naturally express universally quantified conjectures about computable functions.

Interactive proof assistants have long contained functional languages as fragments of the language they support.
  Isabelle/HOL \cite{wenzel2013reference,nipkow2014semantics} offers definitional facilities for functional programming, e.g.\ the \isacommand{datatype} command for inductive data types and \isacommand{fun} for recursive functions.
  A notable feature of Isabelle is its code generator:
certain executable specifications can be translated into source code in target languages such as ML, Haskell, Scala, OCaml \cite{haftmann2009code,haftmann2010code}.
Yet many Scala users do not know Isabelle today.

Aiming to bring the value of trustworthy formalized knowledge to many programmers familiar with Scala, we introduce 
a mapping in the opposite direction: instead of generating code from logic, we show how to map
programs in the purely functional fragment of Scala supported by Leon into Isabelle/HOL.
We use Isabelle's built-in tactics to discharge the verification conditions.
Compared to use of automated solvers in Leon alone, the connection with Isabelle has two main advantages:
  \begin{enumerate}
    \item
      Proofs in Isabelle, even those generated from automated tactics, are justified by a minimal inference kernel.
      In contrast to ATPs, which are complex pieces of software, it is far less likely that a kernel-certified proof is unsound.
    \item
      Isabelle's premier logic, HOL, has seen decades of development of rich mathematical libraries and formalizations such 
      as \href{http://afp.sourceforge.net/}{Archive of Formal Proofs}.
      Proofs carried out in Isabelle have access to this knowledge, which means that there is a greater potential for reuse of existing developments.
  \end{enumerate}
  Establishing the formal correspondence means embedding Scala in HOL, requiring non-trivial transformations (\S\ref{sec:translation}).
  We use a \emph{shallow embedding}, that is, we do not model Scala's syntax, but rather perform a syntactic mapping from Scala constructs to their equivalents in HOL.
  For our implementation we developed an idiomatic Scala API for Isabelle based on previous work by Wenzel \cite{wenzel2014async,wenzel2011isabelle-doc} (\S\ref{sec:tech}).
  We implemented as much functionality as possible inside Isabelle to leverage checking by Isabelle's proof kernel.
  The power of Isabelle's tactics allows us to prove more conditions than what is possible with the Z3 and CVC4 backends (\S\ref{sec:example}).
  We are able to import Leon's standard library and a large amount of its example code base into Isabelle (\S\ref{sec:evaluation}),
  and verify many of the underlying properties.

  \paragraph{Contribution}
  We contribute a mechanism to import functional Scala code into Isabelle, featuring facilities for embedding Isabelle/Isar syntax into Scala via Leon and reusing existing constants in the HOL library without compromising soundness.
  This makes Isabelle available to Leon as a drop-in replacement for Z3 or CVC4 to discharge verification conditions.
  We show that Isabelle automation is already useful for processing such conditions.

  Among related works we highlight a Haskell importer for Isabelle \cite{haftmann2010haskell}, which also uses a shallow embedding and has a custom parser for Haskell, but does not perform any verification.
  Breitner et.\ al.\ have formalised ``large parts of Haskell's standard prelude'' in Isabelle \cite{breitner2013hlints}.
  They use the HOLCF logic, which is extension on HOL for domain theory, and have translated library functions manually.
  Mehnert \cite{mehnert2011kopitiam} implemented a verification system for Java in Coq using separation logic.

  In the following text, we are using the term ``{\purescala}'' to refer to the fragment of Scala supported by Leon \cite[\S 3]{blanc2013overview}, whereas ``Leon'' denotes the system itself.
  More information about Leon and {\purescala} is available from the web deployment of Leon at \url{http://leon.epfl.ch} in the Documentation section.

\section{Bridging the gap}
\label{sec:translation}

  \begin{figure}[t]
    \begin{subfigure}[b]{\linewidth}
      \begin{lstlisting}[language=Scala,gobble=8]
        sealed abstract class List[A]
        case class Cons[A](head: A, tail: List[A]) extends List[A]
        case class Nil[A]() extends List[A]

        def size[A](l: List[A]): BigInt = (l match {
          case Nil => BigInt(0)
          case Cons(_, xs) => 1 + size(xs)
        }) ensuring(_ >= 0)
      \end{lstlisting}
      \caption{Pure Scala version}
      \label{fig:comparison:scala}
    \end{subfigure}
    \begin{subfigure}{\linewidth}
      \begin{lstlisting}[language=Isar,gobble=8]
        datatype 'a list = Nil | Cons 'a "'a list"

        fun size :: "'a list => int" where
        "size Nil = 0" |
        "size (Cons _ xs) = 1 + size xs"

        lemma "size xs >= 0" by (induct xs) auto
      \end{lstlisting}
      \caption{Isabelle version}
      \label{fig:comparison:isabelle}
    \end{subfigure}

    \caption{Example programs: Linked lists and a size function}
    \label{fig:comparison}
  \end{figure}

  Isabelle is a general specification and proof toolkit with the ability of functional programming in its logic Isabelle/HOL.
  Properties of programs need to be stated and proved explicitly in an interactive IDE.
  While the system offers \emph{proof tactics}, the order in which they are called and their parameters need to be specified by the user.
  Users can also write custom tactics which deal with specific classes of problems.

  Leon is more specialised to verification of functional programs and runs in batch mode.
  The user annotates a program and then calls Leon which attempts to discharge resulting verification conditions using ATPs.
  If that fails, the user has to restructure the program.
  Leon has been originally designed to be fully automatic; consequently, there is little support for explicitly guiding the prover.
  However, because of its specialisation, it can leverage more automation in proofs and counterexample finding on first-order recursive functions.

  Due to their differences, both systems have unique strengths.
  Their connection allows users to benefit from this complementarity.

\subsection{Language differences}

  Both languages use different styles in how functional programs are expressed.
  Figure~\ref{fig:comparison} shows a direct comparison of a simple program accompanied by a (trivial) proof illustrating the major differences:
  \begin{itemize}
    \item
      {\purescala} uses an object-oriented encoding of algebraic data types (\emph{sealed clas\-ses}~\cite{odersky2010scala}), similar to Java or C\#.
      Isabelle/HOL follows the ML tradition by having direct syntax support \cite{blanchette2014datatypes}.
    \item
      (Pre-) and postconditions in Leon are annotated using the \verb~ensuring~ function, where\-as Isabelle has a separate \verb~lemma~ command.
      In a sense, verification conditions in Leon are ``inherent'', but need to be stated manually in Isabelle.
    \item
      {\purescala} does not support top-level pattern matching (e.g.\ $\mathit{rev}\;(x\!:\!\mathit{xs}) = \ldots$).
  \end{itemize}
  The translation of data types and terms is not particularly interesting because it is mostly a cavalcade of technicalities and corner cases.
  However, translating functions and handling recursion poses some interesting theoretical challenges.

\subsection{Translating functions}

  A \emph{theory} is an Isabelle/Isar source file comprising a sequence of definitions and proofs, roughly corresponding to the notion of a ``module'' in other languages.
  Theory developments are strictly monotonic.
  Cyclic dependencies between definitions are not allowed \cite{kuncar2015cyclicity}, however, a definition may consist of multiple constants.
  In {\purescala}, there are no restrictions on definition order and cyclicity.

  Consequentially, the Isabelle integration has to first compute the dependency graph of the functions and along with it the set of strongly connected components.
  A single component contains a set of mutually-recursive functions.
  Collapsing the components in the graph then results in a directed acyclic graph which can be processed in any topological ordering.

  The resulting function definitions are not in idiomatic Isabelle/HOL style; in particular, they are not useful for automated tactics.
  Consider Figure \ref{fig:comparison}:
  the naive translation would produce a definition $\mathit{size}\;\mathit{xs} = \mathtt{case}\;\mathit{xs}\;\mathtt{of}\;y\,\#\,ys \rightarrow \ldots \mathit{size}\;\mathit{ys} \ldots$
  Isabelle offers a generic term rewriting tactic (the \emph{simplifier}), which is able to substitute equational rules.
  Such a rule, however, constitutes a non-terminating simplification chain, because the right-hand side contains a subterm which matches the left-hand side.

  This can be avoided by splitting the resulting definition into cases that use Haskell-style top-level pattern matching.
  A verified routine to perform this translation is integrated into Isabelle, producing terminating equations which can be used by automated tactics.
  From this, we also obtain a better induction principle which can be used in subsequent proofs.

  When looking at the results of this procedure, the example in Figure \ref{fig:comparison} is close to reality.
  The given {\purescala} input program produces almost exactly the Isabelle theory below, modulo renaming.
  Because of our implementation, the user normally does not see the resulting theory file (see \S\ref{sec:tech}).
  However, for this example, the internal constructions we perform are roughly equivalent to what Isabelle/Isar would perform (see \S\ref{sec:evaluation}).

\subsection{Recursion}

  Leon has a separate termination checking pass, which can run along with verification and can be turned off.
  Leon's verification results are only meant to be valid under the assumption that its termination checker succeeded (i.e.\ ensuring partial correctness).

  Isabelle's proof kernel does not accept recursive definitions at all.
  We use the \emph{function package} by Krauss \cite{krauss2009function} to translate a set of recursive equations into a low-level, non-recursive definition.
  To automate this construction, the package provides a \isacommand{fun} command which can be used in regular theories (see Figure~\ref{fig:comparison}), but also programmatically.
  To justify its internal construction against the kernel, it needs to prove termination.
  By default, it searches for a lexicographic ordering involving some subset of the function arguments.

  This also means that when Leon is run using Isabelle, termination checking is no longer independent of verification, but rather ``built in''.
  Krauss' package also supports user-specified termination proofs.
  In the future, we would like to give users the ability to write those in Scala.

  A further issue is recursion in data types.
  Negative recursion can lead to unsoundness, e.g.\ introducing non-termination in non-recursive expressions.
  While Leon has not implemented a wellformedness check yet, Isabelle correctly rejects such data type definitions.
  Because we map Scala data types syntactically, we obtain this check for free when using Isabelle in Leon.

\subsection{Cross-language references}

  One of the main reasons why we chose a shallow embedding of {\purescala} into Isabelle is the prospect of reusability of Isabelle theories in proofs of imported {\purescala} programs.
  For example, the dominant collection data structure in functional programming -- and by extension both in {\purescala} and Isabelle/HOL -- are lists.
  Both languages offer dozens of library functions such as \verb~map~, \verb~take~ or \verb~drop~.
  Isabelle's \verb~List~ theory also contains a wealth of theorems over these functions.
  All of the existing theorems can be used by Isabelle's automated tactics to aid in subsequent proofs, and are typically unfolded automatically by the simplifier.

  However, when importing {\purescala} programs, all its data types and functions are defined again in a runtime Isabelle theory.
  While the imported \verb~List.map~ function may end up having the same shape as HOL's \verb~List.map~ function, they are nonetheless distinct constants, rendering pre-existing theorems unusable.

  The naive approach of annotating {\purescala}'s \verb~map~ function to not be imported and instead be replaced by HOL's \verb~map~ function is unsatisfactory:
  The user would need to be trusted to correctly annotate {\purescala}'s library, negatively impacting correctness.
  Hence, we implemented a hybrid approach:
  We first import the whole program unchanged, creating fresh constants.
  Later, for each annotated function, we try to prove an equivalence of the form $f' = f$ where $f'$ is the imported definition and $f$ is the existing Isabelle library function, and register the resulting theorem with Isabelle's automated tools.
  This establishes a trustworthy relationship between the imported {\purescala} program and the existing Isabelle libraries.

  Depending on the size of the analysed program (including dependencies), this approach turns out to be rather inefficient.%
  \footnote{Because our implementation uses Isabelle in interactive instead of in batch mode, we cannot produce pre-computed heap images to be loaded for later runs.}
  According to Leon conventions, we introduced a flag which skips the equivalence proofs for {\purescala} library functions and just asserts the theorems as axioms.
  This also alleviates another practical problem: not all desired equivalences can be proven automatically by Isabelle.
  Support for specifying manual equivalence proofs would be useful, but is not yet implemented.

\section{Technical considerations}
\label{sec:tech}

  Isabelle has been smoothly integrated into Leon by providing an appropriate instance of a \emph{solver}.
  In that sense, Isabelle acts as ``yet another backend'' which is able to check validity of a set of assertions.

\subsection{Leon integration}

  A solver in Leon terminology is a function checking the consistency of a set of assumptions.
  A pseudo-code type signature could be given as $\mathcal P(\mathcal F) \to \{ \mathtt{sat}, \allowbreak \mathtt{unsat}, \allowbreak \mathtt{unknown} \}$, where $\mathcal F$ is the set of supported formulas.
  According to program verification convention, a result of \texttt{unsat} means that no contradiction could be derived from the assumptions, i.e.\ that the underlying program is correct.
  If a solver however returns \texttt{sat}, it is expected to produce a counterexample which violates verification conditions, e.g.\ a value which is not matched by any clause in a pattern match.

  The Isabelle integration is exactly such a function, but with the restriction that it never returns \texttt{sat}, because a failed proof attempt does not produce a suitable counterexample.
  Since Leon offers a sound and complete counterexample procedure for higher-order functions \cite{voirol2015higherorder}, implementing this feature for Isabelle would not be useful.
  
\subsection{Process communication}

  Communication between the JVM process running Leon and the Isabelle process works via our \emph{libisabelle} library which extends Wenzel's PIDE framework \cite{wenzel2014async,wenzel2012isabelle} to cater to non-IDE applications.
  It introduces a remote procedure call layer on top of PIDE, reusing much of its functionality.
  Leon is then able to update and query state stored in the prover process.
  Procedure calls are typed and asynchronous, using an implementation of type classes in ML and Scala's \emph{future} values by Haller et al.\ \cite{haller2012futures}, respectively.

  While being a technologically more complicated approach, it offers benefits over textual Isabelle/Isar source generation.
  Most importantly, because communication is typed, the implementation is much more robust.
  Common sources of errors, e.g.\ pretty printing of Isabelle terms or escaping, are completely eliminated.

\section{Example}
\label{sec:example}

  \begin{figure}[t]
    \begin{lstlisting}[gobble=6,language=Scala]
      def sumReverse[A](xs: List[Nat]) =
        (listSum(xs) == listSum(xs.reverse)).holds
    \end{lstlisting}
    \vskip-5pt
    \begin{lstlisting}[gobble=6,language=Scala]
      @proof(method = """(induct "<var xs>", auto)""")
      def sumConstant[A](xs: List[A], k: Nat) =
        (listSum(xs.map(_ => k)) == length(xs) * k).holds
    \end{lstlisting}
    \vskip-5pt
    \begin{lstlisting}[gobble=6,language=Scala]
      @proof(method = "(clarsimp, induct rule: list_induct2, auto)")
      def mapFstZip[A, B](xs: List[A], ys: List[B]) = {
        require(length(xs) == length(ys))
        xs.zip(ys).map(_._1)
      } ensuring { _ == xs }
    \end{lstlisting}
    \caption{Various induction proofs about lists}
    \label{fig:example}
  \end{figure}

  Figure \ref{fig:example} shows a fully-fledged example of an annotated {\purescala} program.
  As background, assume the \verb~List~ definition from the previous example enriched with some standard library functions, a \verb~Nat~ type, and a \verb~listSum~ function.\footnote{The full example is available at \url{https://git.io/vznVH}.}
  The functions in the example are turned into lemma statements in Isabelle.
  The string parameter of the \verb~proof~ annotation is an actual Isar method invocation, that is, it is interpreted by the Isabelle system.
  For hygienic purposes, names of {\purescala} identifiers are not preserved during translation, but suffixed with unique numbers.
  To allow users to refer back to syntactic entities using their original names, the \verb~<var _>~ syntax has been introduced.

  Running Leon with the Isabelle solver on this example will show that all conditions hold.
  The first proof merely reuses a lemma which is already in the library.
  The other two need specific guidance, i.e.\ an annotation, for them to be accepted by the system.
  The proofs involve Isabelle library theorems, for example distributivity of $(+, *)$ on natural numbers.
  For comparison, Leon+Z3 cannot prove any proposition.
  When also instructed to perform induction, it can prove \verb~sumConstant~.
  (Same holds for Leon+CVC4.)
  There is currently no way in Leon to concisely specify the use of a custom induction rule for Z3 (or CVC4) as required by the last proposition (simultaneous induction over two lists of equal length).

  This example also demonstrates another instance of the general Isabelle philosophy of \emph{nested languages:}
  {\purescala} identifiers may appear inside Isar text which appears inside {\purescala} code.
  Further nesting is possible because Isabelle text can itself contain nested elements (e.g.\ ML code, ...).

\section{Evaluation}
\label{sec:evaluation}

  In this section, we discuss implementation coverage of {\purescala}'s syntactic constructs, trustworthiness of the translation and overall performance.

\paragraph{Coverage.}
  The coverage of the translation is almost complete.
  A small number of Leon primitives, among them array operations %
  have not been implemented yet.\footnote{In fact, while
    attempting to implement array support we
  discovered that Leon's purely functional view of immutably used arrays 
  does not respect Scala's reference equality implementation of arrays, leading
  to a decision to disallow array equality in Leon's {\purescala}.}
  All other primitives are mapped as closely as possible and adaptations to Isabelle are proven correct when needed.
  Leon's standard library contains -- as of writing -- 177 functions with a total of 289 verification conditions, out of which Isabelle can prove 206 ($\approx$ 71\%).

\paragraph{Trustworthiness.}
  Our mapping uses only definitional constructs of Isabelle and thus
  the theorems it proves have high degree of trustworthiness.  
  Using a shallow embedding always carries the risk of semantics mismatches.
  A concern is that since the translation of {\purescala} to Isabelle works through an internal API, the user has no possibility to convince themselves of the correctness of the implemented routines short of inspecting the source code.
  For that reason, all operations are logged in Isabelle.
  A user can request a textual output of an Isar theory file corresponding to the imported {\purescala} program, containing all definitions and lemma statements, but no proofs.
  This file can be inspected manually and re-used for other purposes,
  and represents faithfully the facts that Isabelle actually proved in
  a readable form.

\paragraph{Performance.}
  On a contemporary dual-core laptop, just defining all data types from the {\purescala} library (as of writing: 13), but no functions or proofs, Leon+Isabelle takes approximately 30 seconds.
  Defining all functions adds another 70 seconds to the process.
  Using Leon+Z3, this is much faster: it takes less than 10 seconds.
  The considerable difference (factor $\approx$ 10) can be explained by looking at the internals of the different backends.
  Z3 has data types and function definitions built into its logic.
  Isabelle itself does not: both concepts are implemented in HOL, meaning that every definition needs to be constructed and justified against the proof kernel.
  The processing time of an imported {\purescala} programs is comparable to that of a hand-written, idiomatic Isabelle theory file.
  In fact, during processing the {\purescala} libraries, thousands of messages are passed between the JVM and the Isabelle process, but the incurred overhead is negligible compared to the internal definitional constructions.

\section{Conclusion}

  We have implemented an extension to Leon which allows using Isabelle to discharge verification conditions of {\purescala} programs.
  Because it supports the vast majority of syntax supported by Leon, we consider it to be generally usable.
  It is incorporated in the Leon source repository,\footnote{\url{https://github.com/epfl-lara/leon}} supporting the latest Isabelle version (Isabelle2016).

  With this work, it becomes possible to co-develop a specification in both {\purescala} and Isabelle, use Leon to establish a formal correspondence, and prove interesting results in Leon and/or Isabelle/Isar.
  Because of the embedded Isar syntax, complicated correctness proofs can also be expressed concisely in Leon.
  To the best of our knowledge, this constitutes the first bi-directional integration between a widespread general purpose programming language and an interactive proof assistant.

  An unintended consequence is that since Isabelle can export code in Haskell and now import code from {\purescala}, there is a fully-working Scala-to-Haskell cross-compilation pipeline.
  The transformations applied to the {\purescala} code to make it palatable to Isabelle's automation also results in moderately readable Haskell code.

\paragraph{Acknowledgements}
  We would like to thank the people who helped ``making the code work'': Ravi Kandhadai, Etienne Kneuss, Manos Koukoutos, Mikäel Mayer, Nicolas Voirol, Makarius Wenzel.
  Cornelius Diekmann, Manuel Eberl, and Tobias Nipkow suggested many textual improvements to this paper.

\bibliography{references}
\bibliographystyle{splncs03}

\end{document}